\documentclass[12pt]{article}

\usepackage[utf8]{inputenc}   % ensure UTF-8 input encoding
\usepackage[T1]{fontenc}      % good font encoding for PDF output

\usepackage[margin=1in]{geometry}
\usepackage{setspace}
\usepackage{amsmath,amssymb}
\usepackage{graphicx}
\usepackage{booktabs}
\usepackage{longtable}
\usepackage{float}
\usepackage{placeins}
\usepackage{siunitx}
\usepackage{caption}
\usepackage{threeparttable}
\usepackage{makecell}
\usepackage[numbers,sort&compress]{natbib} % numeric citation, compressed ranges
\usepackage{url}                            % safe URLs in bibliography

\usepackage{courier}          % sets \ttdefault to 'pcr' (Courier)

\newcommand{\itsa}{\texttt{itsa}}  % use \itsa{} in text to typeset in Courier

\usepackage{hyperref}
\usepackage{doi}

\hypersetup{
  colorlinks=true,
  linkcolor=blue,
  citecolor=blue,
  urlcolor=blue
}
\title{\textbf{Improving causal inference in interrupted time series analysis: the triple difference design}}
\author{
Ariel Linden, DrPH \\
University of California, San Francisco \\
Department of Medicine \\
Division of Clinical Informatics \& Digital Transformation (DoC-IT) \\
ariel.linden@ucsf.edu
}
\date{}

\begin{document}

\maketitle

\begin{abstract}
Background: Interrupted time series analysis (ITSA) is a valuable quasi-experimental design for evaluating health policy and intervention effects. While multiple-group ITSA (MG-ITSA) can improve causal inference under additional identifying assumptions by incorporating a control group, residual confounding from unmeasured time-varying factors may persist. The triple-difference interrupted time series (DDD-ITSA) design extends this framework by adding a second control group to further isolate treatment effects, yet the design remains underutilized and lacks formal methodological guidance.

Methods: This paper formalizes the DDD-ITSA design for health research. We specify the underlying regression model, define the key coefficients for estimating level and trend effects, and clarify the interpretation of the triple-difference estimand. We also illustrate application of the design using a worked example evaluating the effect of California’s Proposition 99 cigarette tax increase on per-capita cigarette sales.

Results: The DDD-ITSA model provides clear parameterization for assessing baseline balance across groups and estimating differential treatment effects. In the stylized Proposition 99 example, all groups demonstrated balance on pre-intervention level and trend. The triple-difference estimand revealed a statistically significant annual net reduction of -1.76 per-capita cigarette packs in California relative to the secondary control (P = 0.020; 95\% CI: -3.24, -0.28), consistent with the treatment effect observed against the primary control. The difference-in-differences between the two control groups was non-significant, supporting the validity of the comparison.

Conclusions: The DDD-ITSA design offers a useful extension of the interrupted time series framework that can improve causal inference when its identifying assumptions are plausible. By leveraging an additional control group, researchers can difference out remaining biases and assess effect heterogeneity across groups or comparison dimensions. The updated \itsa{} package for Stata facilitates implementation, making this design accessible for routine application in health policy research. Careful attention to control group selection, baseline balance, and autocorrelation remains essential for valid interpretation.
\end{abstract}

\section*{Keywords}
triple differences, interrupted time series, difference-in-differences, causal inference, quasi-experimental design, health policy

% -------------------------
% Background
% -------------------------
\section{Background}

Interrupted time series analysis (ITSA) is widely used in healthcare research to estimate the causal effects of interventions, policy changes, and natural events when randomized designs are infeasible. The approach relies on sequential outcome measurements collected at equally spaced intervals, typically aggregated at the unit level (e.g., hospitals, regions, or populations) and summarized as rates, counts, or mean expenditures. In its canonical form, a single unit is exposed to an intervention that is hypothesized to generate a structural change (“interruption”) in the level and/or slope of the outcome process at a known point in time \cite{campbell1966,shadish2002}. Identification in this framework is achieved by using the pre-intervention segment of the series to extrapolate the counterfactual post-intervention trajectory under the assumption that, absent the intervention, the underlying data-generating process would have continued unchanged. The use of multiple pre-intervention observations partially addresses threats such as regression to the mean and improves internal validity relative to simpler before–after comparisons \cite{campbell1966,shadish2002,linden2007}.

Despite these strengths, single-group ITSA (SG-ITSA) rests on the strong identifying assumption that no unmeasured, time-varying confounders or coincident events affect the outcome at the time of intervention. When this assumption is violated (such as in the presence of secular trends, anticipatory behavior, or concurrent policy changes), SG-ITSA may produce biased estimates of the intervention effect. Empirical evidence indicates that SG-ITSA can yield misleading inferences when pre-intervention trends already exhibit directional change or when external influences evolve contemporaneously with the intervention \cite{linden2016,linden2017a,linden2017b}.

Multiple-group or controlled ITSA (MG-ITSA) addresses these threats by introducing control units that were not exposed to the intervention but share similar baseline levels, pre-intervention trends, and observed characteristics with the treated unit \cite{linden2015,abadie2010,linden2018}. By differencing outcomes between treated and control series over time, MG-ITSA relaxes the strong assumption of stable counterfactual trends within a single unit and instead relies on a parallel-trends-type assumption across groups \cite{linden2015,linden2018}. Under this framework, causal effects are identified by removing secular trends and common shocks that affect all units similarly, aligning MG-ITSA with difference-in-differences estimators, which it generalizes to settings with multiple time points and allows for estimation of both level and trend changes. However, when researchers have access to multiple plausible comparison groups and are concerned that a single control group may not adequately account for unmeasured time-varying confounding, standard MG-ITSA may be insufficient to isolate the causal effect of interest.

A triple-difference interrupted time series (DDD-ITSA) design extends the logic of the MG-ITSA by adding a second control group for comparison. This approach mirrors the difference-in-differences-in-differences (DDD) framework in the econometrics literature, where an additional difference is used to remove remaining confounding that may persist after a standard two-group comparison \cite{olden2022}. In the ITSA context, the triple-difference estimator captures whether the post-intervention change in level and/or trend for the treated unit differs relative to an additional control-group comparison beyond what would be expected based on the primary treatment--control comparison. By differencing across time, groups, and an additional comparison dimension, the DDD-ITSA design can improve causal inference under assumptions analogous to those in the difference-in-differences (DiD) literature, namely that, absent the intervention, these relative differences would have evolved similarly over time \cite{ryan2019}.

Applications that explicitly implement a DDD-ITSA framework remain relatively uncommon in the literature. One applied example is the evaluation of Maryland’s global budget revenue (GBR) model, where researchers used an interrupted time-series design with difference-in-differences comparisons and a triple differences analysis to assess changes in emergency department utilization, admissions, and revisits following GBR implementation, including how these trends varied across racial/ethnic and payer groups \cite{galarraga2022}. In this study, the triple-difference component helped identify differential effects of the policy on emergency department return outcomes across subpopulations beyond general statewide trends and group-level effects. Another example is research on the effects of the Affordable Care Act (ACA) on health insurance coverage among Latino subgroups, which combined ITSA with triple-difference models to evaluate heterogeneous changes in uninsured rates associated with the ACA and Medicaid expansion across multiple Latino ethnic groups, highlighting substantial variation in coverage gains that would not be captured by simpler two-group comparisons \cite{gonzales2018}. More generally, the triple-difference framework identifies differences in treatment effects across groups (i.e., effect heterogeneity), while estimation of the treatment effect within a specific subgroup represents a special case that requires the additional assumption that the comparison group experiences no treatment effect.

This paper formalizes the DDD-ITSA framework for applied health research. We describe the design, detail the underlying statistical model, clarify the interpretation of the key coefficients that identify level and trend effects, and present a worked empirical example to illustrate implementation and interpretation. We also provide practical guidance regarding identification assumptions, model specification, and appropriate use cases for the design. To facilitate applied use, the \itsa{} package for Stata \cite{linden2015} has been updated to estimate DDD-ITSA models, enabling researchers to easily implement the approach using their data.

% -------------------------
% Methods
% -------------------------
\section{Methods}

\subsection{The DDD-ITSA design}

\begin{figure}[H]
\centering
\includegraphics[width=\textwidth]{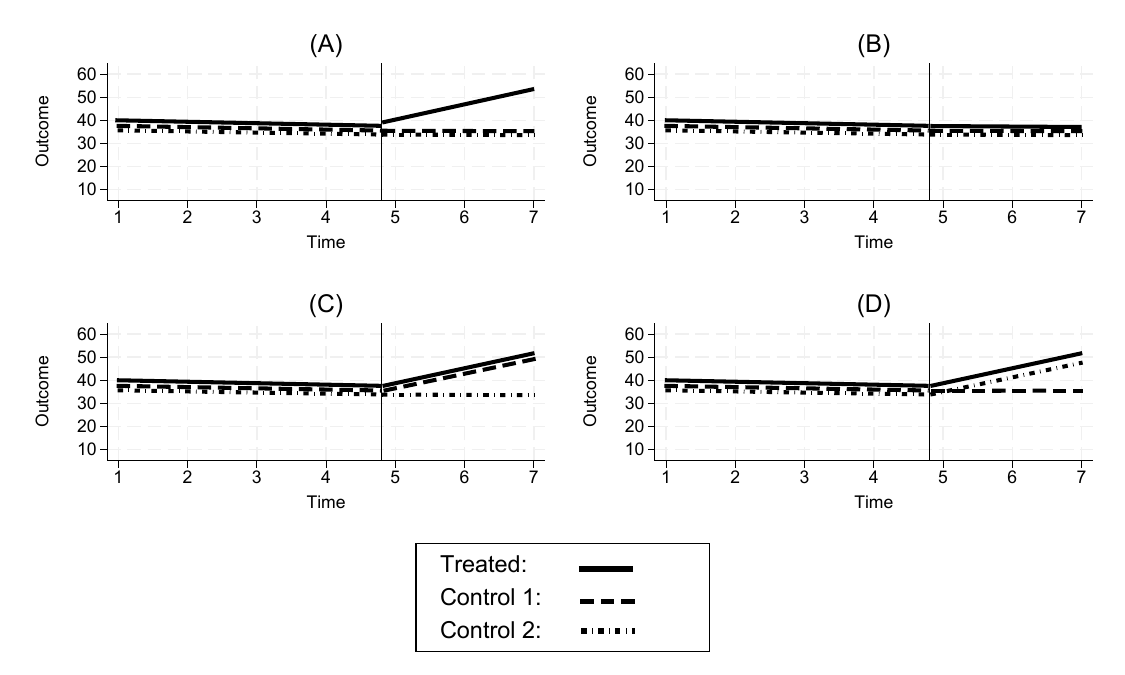}
\caption{Four general outcome scenarios for the DDD-ITSA. (A) A treatment effect in the primary analysis, (B) No treatment effect, (C) A treatment effect in the secondary analysis (evidence of confounding in the primary comparison), (D) A treatment effect in the primary analysis, but evidence of confounding in the secondary analysis.}
\label{fig:scenarios}
\end{figure}

The DDD-ITSA design is an extension of the MG-ITSA design to an additional control group that is either geographically distant from the treatment and primary control groups, or comprises a different demographic that still allows for comparison (e.g., type of insurance coverage, different medical condition, ethnicity, etc.). This second control group provides an additional comparison that can help assess and potentially account for residual confounding by evaluating whether biases in the primary comparison are shared across groups. Figure~\ref{fig:scenarios} illustrates four of the many potential outcome scenarios that we could anticipate in the DDD-ITSA design when emphasizing differences in trends: (A) a treatment effect in the primary analysis and secondary analysis (the treatment unit's post treatment trend increases, while the primary control group and secondary control groups' post-treatment trends remain consistent with how they were trending in the pre-treatment period); (B) no treatment effect in either the primary or secondary analysis (all groups' post-treatment trends continue trending consistently with how they were trending in the pre-treatment period); (C) a treatment effect in the secondary analysis but not in the primary analysis (there is evidence of confounding either by spill-over or by an external event); and (D) a treatment effect in the primary analysis but not in the secondary analysis (suggesting that something outside the intervention is influencing outcomes in the secondary control group such as an economic trend, seasonal effect, or spillover from a neighboring intervention). While Figure~\ref{fig:scenarios} focuses on stylized outcome scenarios, it also provides intuition for the underlying triple-difference logic by illustrating how comparisons across time, groups, and an additional comparison dimension can reveal whether observed changes are attributable to the intervention or to confounding influences.

\FloatBarrier

\subsection{Identification assumptions}

The validity of the DDD-ITSA estimator rests on a set of identifying assumptions that extend the logic of difference-in-differences estimators \cite{olden2022}. Specifically, identification requires that, in the absence of the intervention, the difference between the two-component difference-in-differences comparisons would have remained constant over time. This condition can be understood as a generalized parallel trends assumption \cite{ryan2019}. Let $Y_{gt}^{0}$ denote the potential outcome under no intervention for group $g$ at time $t$. The identifying assumption can then be written as:

\begin{equation}
\begin{split}
&\left[E\left(Y_{\mathit{Tx},t}^{0} - Y_{\mathit{C1},t}^{0}\right) - E\left(Y_{\mathit{Tx},t-1}^{0} - Y_{\mathit{C1},t-1}^{0}\right)\right] \\
&\qquad = \left[E\left(Y_{\mathit{C2},t}^{0} - Y_{\mathit{C1},t}^{0}\right) - E\left(Y_{\mathit{C2},t-1}^{0} - Y_{\mathit{C1},t-1}^{0}\right)\right]
\end{split}
\label{eq:ident}
\end{equation}

\noindent where $\mathit{Tx}$ denotes the treated unit, $\mathit{C1}$ the primary control group, and $\mathit{C2}$ the secondary control group. This condition states that, in the absence of the intervention, the difference-in-differences comparison between the treated unit and the primary control group would have evolved in the same way as the difference-in-differences comparison between the secondary and primary control groups. While the DDD estimator is sometimes described as requiring only a single parallel trends condition, in practice it implies that any bias affecting the primary comparison between the treatment and first control group must also affect the comparison between the second control group and the first control group in an equivalent way. Under this condition, the shared bias component is differenced out, yielding an unbiased estimate of the treatment effect \cite{olden2022}. If the sources of non-parallel trends differ across these comparisons, then the triple-difference estimator may remain biased. As such, careful selection of control groups and substantive justification of the identifying assumptions are critical for valid causal interpretation. Conceptually, the DDD-ITSA estimator can be understood as the difference between two difference-in-differences comparisons, with the additional comparison dimension providing a second contrast that allows shared biases to be differenced out. Although this representation implies that one of the control-group contrasts cancels algebraically from the simplified expression of the estimand, this reflects a property of the parameterization rather than the absence of the control groups' role in identification. Consistent with the triple-difference framework described by Olden and Møen \cite{olden2022}, causal interpretation depends on the assumption that differential trends across the treated and control comparisons would have evolved similarly in the absence of the intervention. In the present framework, the primary control group provides the standard counterfactual comparison for the treated unit, while the secondary control group provides an additional comparison that helps assess whether biases in the primary comparison are shared and can therefore be differenced out.

\subsection{The DDD-ITSA model}

The DDD-ITSA regression model is an extension of the MG-ITSA \cite{linden2015,linden2017c,linden2022erratum} and assumes the following form:

\begin{equation}
\begin{aligned}
Y_t &= \beta_0 + \beta_1 T_t + \beta_2 X_t + \beta_3 (X_t \times T_t) + \beta_4 Z_1 + \beta_5 (Z_1 \times T_t) \\
    &\quad + \beta_6 (Z_1 \times X_t) + \beta_7 (Z_1 \times X_t \times T_t) + \beta_8 Z_2 + \beta_9 (Z_2 \times T_t) \\
    &\quad + \beta_{10} (Z_2 \times X_t) + \beta_{11} (Z_2 \times X_t \times T_t) + \epsilon_t
\end{aligned}
\label{eq:model}
\end{equation}

\noindent where $Y_t$ is the aggregated outcome variable measured at each time-point $t$, $T_t$ is the time since the start of the study; $X_t$ is a dummy variable representing the treatment (pre-treatment periods = 0, otherwise 1); $Z_1$ is a dummy variable to denote the treatment unit and $Z_2$ is a dummy variable to denote the second control group. Therefore, all variables and coefficients that include $Z_1$ are equivalent to those in an MG-ITSA model \cite{linden2015} in which the treatment unit is compared to the control group, and all variables and coefficients that include $Z_2$ compare the second control group to the first control group. More specifically, $\beta_4$ represents the difference in the level (intercept) of the dependent variable between treatment and the first control group prior to the intervention, $\beta_5$ represents the difference in the slope (trend) of the dependent variable between treatment and the first control group prior to the intervention, $\beta_6$ indicates the difference between treatment and the first control group in the level of the dependent variable immediately following introduction of the intervention, and $\beta_7$ represents the difference between treatment and the first control group in the slope (trend) of the dependent variable after initiation of the intervention compared with pre-intervention (akin to a difference-in-differences of slopes). $\beta_8$ represents the difference in the level (intercept) of the dependent variable between the second control group and the first control group prior to the intervention, $\beta_9$ represents the difference in the slope (trend) of the dependent variable between the second control and the first control group prior to the intervention, $\beta_{10}$ indicates the difference between the second control group and the first control group in the level of the dependent variable immediately following introduction of the intervention, and $\beta_{11}$ represents the difference between the second control group and the first control group in the slope (trend) of the dependent variable after initiation of the intervention compared with pre-intervention.

Although Equation~\eqref{eq:model} provides a convenient parameterization of group-level trends and intervention effects, it is important to distinguish between the statistical model and the assumptions required for causal identification. The regression specification describes the structure of the outcome over time, including baseline levels, trends, and their changes following the intervention. However, the ability to interpret the estimated coefficients causally does not arise from the functional form of the model itself, but rather from the validity of the identifying assumptions described above. In particular, causal interpretation relies on the comparability of trends across groups and the assumption that any bias affecting the component comparisons is equivalent and can be differenced out. The model therefore serves as a tool for estimation, while identification is grounded in the study design and underlying assumptions.

When the random error terms follow an AR(1) process,

\begin{equation}
\epsilon_t = \rho \epsilon_{t-1} + u_t
\label{eq:ar1}
\end{equation}

\noindent where the autocorrelation parameter $\rho$ is the correlation coefficient between adjacent error terms, such that $|\rho| < 1$, and the disturbances $u_t$ are independent $N(0,\sigma^2)$ (see \cite{kutner2005} for a detailed discussion of autocorrelation in time-series regression models).

Table~\ref{tab:trends} presents the coefficients and linear combination terms for the pre-treatment and post-treatment trends of each of the three groups, and Table~\ref{tab:did_trends} presents the difference-in-differences of trend comparisons between groups. As shown, $(\beta_7 - \beta_{11})$ represents the difference-in-differences in trends between the treated unit and the second control group. Due to the specified parameterization, this expression is algebraically equivalent to the triple-difference estimand, which can also be written as $(\beta_7 - \beta_{11}) = [(\beta_3 + \beta_7) - \beta_3] - [(\beta_3 + \beta_{11}) - \beta_3]$. This equivalence arises from the choice of reference group and the structure of the regression model, which causes the primary control group terms to cancel out. Importantly, this algebraic simplification reflects how the estimand is represented within the model and does not, by itself, justify a causal interpretation. Causal interpretation instead depends on the identifying assumptions described above, particularly the requirement that any bias affecting the component comparisons is equivalent and can be differenced out.

Tables~\ref{tab:levels} and \ref{tab:did_levels} present the analogous coefficients and linear combination terms for the pre-treatment and post-treatment level changes of each of the three groups and their respective difference in differences. The DDD estimate for the immediate level effect (that is, the effect occurring in the period immediately following introduction of the treatment) is given by $(\beta_6 - \beta_{10})$, representing the additional post-intervention level change for the treatment group relative to the second comparison group, net of secular changes observed in the holdout. Together, these parameters quantify both the short-run (level) and longer-run (trend) causal effects of the intervention within a triple-difference interrupted time series framework.

\subsection{Assessing balance on baseline level and trend}

In non-experimental settings, it is critical to evaluate whether the treatment and control groups are balanced in both the level and slope of the outcome variable prior to the intervention. In a randomized controlled trial, the expectation is that baseline levels and trends are statistically equivalent across groups, supporting identification of causal effects. In observational studies, however, pre-treatment differences in intercepts or trends may violate the parallel trends assumption, potentially biasing estimates and complicating causal inference regarding the impact of the intervention on the outcome.

Balance across groups in baseline levels and trends can be formally assessed using the coefficients from Equation~\eqref{eq:model}. Specifically, statistically insignificant estimates of $\beta_4$ and $\beta_5$ indicate no systematic differences between the treatment group and the primary control group in pre-intervention levels and trends, respectively. Likewise, statistical insignificance of the linear combinations $(\beta_4 - \beta_8)$ and $(\beta_5 - \beta_9)$ suggests baseline equivalence between the treatment group and the secondary control group in levels and trends. Finally, statistically insignificant $\beta_8$ and $\beta_9$ imply balance between the two control groups in pre-treatment levels and trajectories.

Although multiple baseline trend comparisons are evaluated, Olden and Møen \cite{olden2022} demonstrate that identification of the DDD estimator does not require two independent parallel trends assumptions. The intuition is that the DDD estimator can be interpreted as the difference between two potentially biased difference-in-differences estimators. As long as the bias component is identical across these estimators, it differences out, yielding an unbiased DDD estimate. Consequently, only a single parallel trends condition is required for causal interpretation.

\subsection{Example}

The following example illustrates the analysis of the DDD-ITSA design using data from California's 1988 enactment of Proposition 99, a voter-approved initiative that increased the cigarette excise tax by \$0.25 per pack and funded statewide anti-smoking efforts. The outcome examined is per-capita cigarette sales (packs), a standard proxy for smoking prevalence in the tobacco control literature, measured annually at the state level from 1970 to 2000, with 1989 marking the first post-intervention year. The dataset, compiled by Abadie et al.\ \cite{abadie2010}, includes cigarette sales and four covariates: average retail cigarette price, logged per-capita personal income, per-capita beer consumption, and the share of the population aged 15--24. Eleven states that implemented major tobacco control programs during 1989--2000 were excluded, leaving 38 states as potential controls.

For exposition purposes, we assign Idaho and Montana to the primary control group and Colorado as the secondary control state. These states were identified as balanced matches to California (the treated state) on baseline level and trend of cigarette sales and retail price using the \texttt{itsamatch} program in Stata \cite{linden2018}. We analyze the model using the new DDD-ITSA feature in the \itsa{} program for Stata \cite{linden2015}. We specify regression with Newey--West standard errors \cite{newey1987} as the statistical model, and specify a single lag autoregressive structure (AR1) based on the results of the \texttt{actest} \cite{baum2013}. The Stata post-estimation command \texttt{lincom} is used to compute the Wald estimate for linear combinations of coefficients. The exact code used in this example is found in the Appendix.
% -------------------------
% Results
% -------------------------
\section{Results}

Table~\ref{tab:regression} presents the regression results for the DDD-ITSA model. First, we review how well the groups are balanced on baseline level and trend. $\beta_4$ and $\beta_5$ indicate that California and the primary control states (Idaho and Montana) are balanced on baseline level and trend ($P = 0.347$ and $P = 0.539$, respectively). A linear combination of terms of $(\beta_4 - \beta_8)$ and $(\beta_5 - \beta_9)$ indicate that California and Colorado (the secondary control state) are also balanced on baseline level and trend ($P = 0.325$ and $P = 0.711$ for level and trend, respectively; data not shown). Finally, $\beta_8$ and $\beta_9$ indicate that the two control groups are also balanced on baseline level and trend ($P = 0.072$ and $P = 0.880$, for level and trend, respectively).

\begin{figure}[H]
\centering
\includegraphics[width=\textwidth]{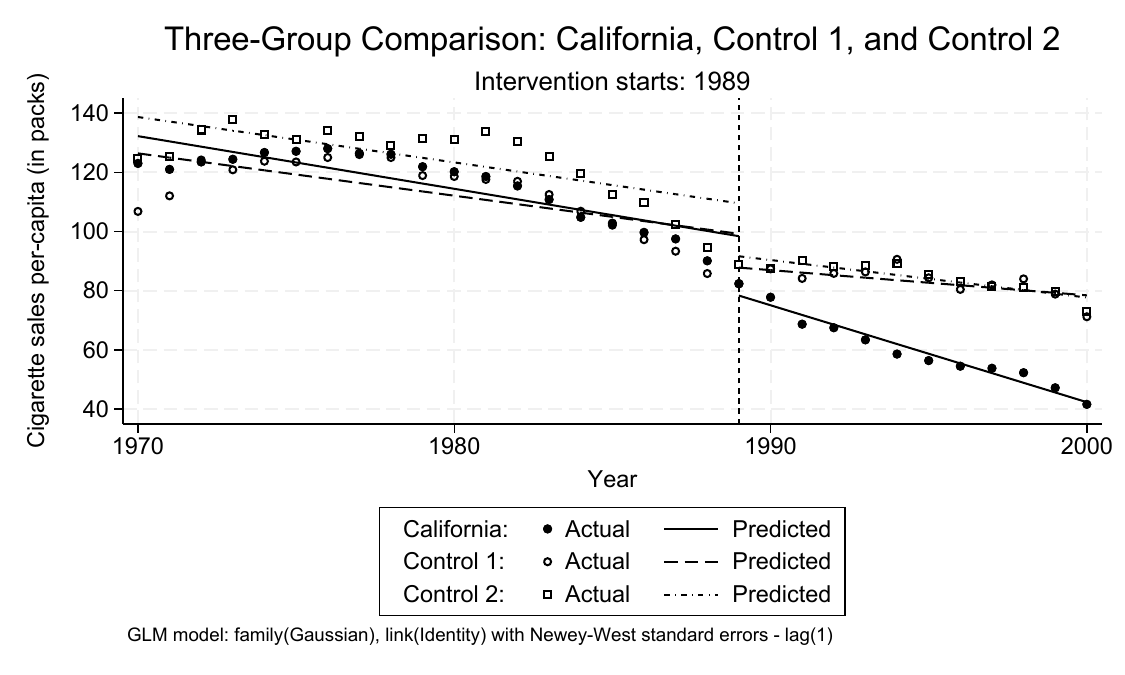}
\caption{Graphic display of the DDD-ITSA outcomes produced by the \itsa{} package in Stata.}
\label{fig:results}
\end{figure}

Next, we evaluate treatment effects amongst the groups, with an emphasis on differences-in-differences in trends. $\beta_7$ indicates that California experienced a statistically significant annual net reduction (post-Proposition 99 -- pre-Proposition 99) of $-2.07$ per-capita cigarette sales (in packs) as compared to the primary control states ($P = 0.006$; 95\% CI: $-3.54$, $-0.61$). Similarly, the linear combination of terms $(\beta_7 - \beta_{11})$ indicates that California had a statistically significant annual net decrease of $-1.76$ per-capita cigarette sales (in packs) as compared to the secondary control ($P = 0.020$; 95\% CI: $-3.24$, $-0.280$). The $\beta_{11}$ coefficient indicates that the difference-in-differences between the primary and secondary control groups was a non-statistically significant annual decrease of $-0.31$ per-capita cigarette sales (in packs) ($P = 0.723$; 95\% CI: $-2.03$, $1.41$). As explained in the Methods section, the overall DDD-ITSA is the same as the treated unit versus the secondary control group, which in this example is a net annual reduction of $-1.76$ per-capita cigarette sales (in packs) ($P = 0.020$; 95\% CI: $-3.24$, $-0.280$). Figure~\ref{fig:results} presents a graphic display of the results that are produced by \itsa{}. Taken together, these results are similar to those presented in Figure~\ref{fig:scenarios}(A), in which a treatment effect is realized against both sets of controls, and no false effects are noted between control groups. This example is intended to illustrate the implementation and interpretation of the DDD-ITSA framework under conditions where identifying assumptions appear plausible. In practice, however, violations of these assumptions (such as differential non-parallel trends or time-varying confounders that affect the component comparisons differently) may lead to biased estimates, underscoring the importance of careful assumption checking in applied analyses.

Visual inspection of Figure~\ref{fig:results} indicates some curvature in the pre-intervention trajectories, such that the linear specification should be regarded as an approximation. In a sensitivity analysis allowing group-specific quadratic time trends, entered as covariates in \itsa{}, the common post-intervention level change ($\beta_2$) changed from $-11.43$ to $4.18$, indicating that the apparent common level reduction was sensitive to the assumed functional form. However, the principal finding of a greater post-intervention decline in the trend of cigarette sales in California relative to the secondary control was retained under the quadratic specification ($-4.45$ packs per capita per year; 95\% CI: $-6.77$, $-2.13$; $P < 0.001$), compared with $-1.76$ under the linear specification. Thus, the qualitative conclusion regarding the differential trend was unchanged, although its magnitude and the estimated level changes were sensitive to specification.

\FloatBarrier

% -------------------------
% Discussion
% -------------------------
\section{Discussion}

This paper formalizes the triple-difference interrupted time series (DDD-ITSA) design, an extension of the multiple-group ITSA framework that incorporates an additional control group to improve causal inference under a complementary set of identifying assumptions in health policy and intervention research. The DDD-ITSA approach combines the structural time-series logic of ITSA with the econometric rigor of difference-in-differences-in-differences estimators, offering researchers a powerful tool for estimating causal effects when standard two-group comparisons may remain confounded by unmeasured time-varying factors or differential trends across groups or comparison dimensions. The key contributions of this paper are the explicit articulation of the DDD-ITSA model, including clear parameter interpretation, and the worked example to illustrate the practical application of the DDD-ITSA design.

Researchers implementing DDD-ITSA face several issues with the design that require careful consideration. First, the choice of primary and secondary control groups is substantively important and should be guided by content expertise and empirical balance assessments. An additional consideration is the potential for spillovers or interference between groups. The validity of the DDD-ITSA design relies on the assumption that the intervention does not directly or indirectly affect outcomes in the control groups. In many policy settings, however, spillover effects may arise through mechanisms such as geographic proximity, behavioral responses, or market interactions. If such interference is present, the estimated treatment effects may be biased. Researchers should therefore carefully consider the plausibility of the no-interference assumption and, where possible, select control groups that are unlikely to be affected by the intervention. The primary control should be as similar as possible to the treatment unit on observed characteristics and pre-treatment trends \cite{linden2015,linden2018}, while the secondary control provides an additional layer of adjustment, ideally one that shares the primary control's susceptibility to common shocks but differs in ways that help isolate treatment effects.

Second, when the pool of potential controls is sufficiently large, increasing the number of units assigned to each control group should be considered. Linden \cite{linden2026} found that adding more control units in the MG-ITSA design reduced standard errors and increased power. By extension, incorporating additional comparison groups in a DDD-ITSA would similarly enhance precision by providing more stable estimates of the counterfactual trends across multiple dimensions.

Third, researchers should carefully assess and account for autocorrelation, as failure to do so can lead to biased standard errors and invalid statistical inference, even when the underlying model is correctly specified. This can be accomplished by reviewing autocorrelation and partial autocorrelation functions \cite{box2016} and formally testing for autocorrelation using appropriate diagnostics (e.g., the \texttt{actest} command in Stata \cite{baum2013}) and specifying error structures accordingly, whether through Newey--West standard errors, Prais--Winsten regression \cite{prais1954}, Cochrane--Orcutt regression \cite{cochrane1949}, instrumental variables \cite{linden2006}, or any one of the autoregressive conditional heteroskedasticity family of estimators (see Harvey \cite{harvey1989} and Enders \cite{enders2004} for a comprehensive discussion of econometric time series models).

Fourth, researchers should endeavor to collect as long of time series data as possible. Because the DDD-ITSA model includes a larger number of parameters (12 in total), it introduces greater variance in the estimated coefficients, which can substantially reduce precision and statistical power relative to SG-ITSA or MG-ITSA designs. As a result, detecting meaningful effects, particularly changes in trends, typically requires longer time series with sufficient pre- and post-intervention observations. Linden \cite{linden2026} found that in the MG-ITSA context, a longer study period provides the necessary statistical resolution to detect trend-based effects and mitigates the bias introduced by autocorrelation, which can otherwise severely undermine power. These considerations are likely to be even more important in DDD-ITSA, where additional parameters increase variance and may reduce precision in shorter series.

Fifth, the functional form of the underlying time trend may not be linear. Researchers should therefore assess the pre-intervention trajectories and use an appropriate specification that adequately represents the observed data.

Finally, Linden \cite{linden2026} found that introducing treatment at the midpoint of the time series maximizes power for identifying trend changes in the MG-ITSA. This would apply equally to DDD-ITSA designs, as balanced pre- and post-treatment periods would minimize standard errors across all interaction terms.

The DDD-ITSA design is particularly well-suited to several common scenarios in health services and policy research. First, when researchers suspect that two-group comparisons may be confounded by concurrent policy changes, economic shocks, or secular trends that differentially affect groups or comparison dimensions, the addition of a second control group can difference out these common sources of bias. Second, DDD-ITSA is valuable when treatment effects are expected to vary across an additional dimension such as demographic groups or comparison dimensions, payer types, or clinical conditions. Third, the design is appropriate when researchers have access to multiple potential control groups but are uncertain which provides the most valid counterfactual. Rather than choosing arbitrarily or averaging across controls, the DDD-ITSA framework leverages both comparisons simultaneously, with the triple-difference estimand providing a summary effect that differences out bias common to the two component comparisons. This interpretation relies on the assumption that any bias affecting the underlying difference-in-differences comparisons operates similarly across them, rather than on the validity of any single comparison in isolation. Fourth, DDD-ITSA can serve as a robustness check for standard MG-ITSA findings. When two-group and three-group analyses yield consistent results, confidence in causal inferences increases. When results diverge, as in scenarios C and D in Figure~\ref{fig:scenarios}, the pattern of findings can illuminate potential threats to validity and guide further investigation. For instance, a treatment effect in the secondary but not primary analysis (scenario C) might suggest spillover effects or confounding by events affecting the primary control, while effects in the primary but not secondary analysis (scenario D) might indicate that the secondary control is subject to unique influences that compromise its validity.

A key identifying assumption underlying the DDD-ITSA framework is that the difference between the two difference-in-differences comparisons would have remained constant in the absence of the intervention. While this is sometimes framed as requiring only a single parallel trends condition, in practice it implies that any bias present in the primary comparison must be replicated in the secondary comparison in order to be differenced out. If the factors generating non-parallel trends differ across these comparisons, the DDD estimate may remain biased. Researchers should therefore carefully consider the plausibility of this assumption when selecting control groups and interpreting and reporting results \cite{linden2005}.

Despite the complexities of the design, the newly added DDD-ITSA functionality in the \itsa{} package for Stata \cite{linden2015} simplifies the implementation and offers many options to ease interpretation and graphically display the outcomes (see the \itsa{} help file for a complete description of options and additional examples). The interested reader will find all of the code used here for computing results of the Proposition 99 example in the Appendix.

% -------------------------
% Conclusion
% -------------------------
\section{Conclusion}

The DDD-ITSA design offers a valuable addition to the health policy researcher's toolkit, extending the logic of multiple-group interrupted time series to address remaining sources of confounding through the inclusion of a second control group. By formalizing the model, clarifying parameter interpretation, and providing practical implementation guidance, this paper aims to facilitate appropriate use of the design. The updated \itsa{} package for Stata makes DDD-ITSA accessible for routine application, enabling researchers to improve causal inference when the identifying assumptions of the design are plausible and supported by substantive knowledge and empirical assessment. As with all quasi-experimental methods, the validity of DDD-ITSA estimates ultimately depends on the plausibility of its identifying assumptions in the specific context of application, underscoring the importance of substantive knowledge, transparent reporting, and critical evaluation of evidence.

\bibliographystyle{unsrtnat}
\bibliography{references}

@book{campbell1966,
  author = {Campbell, Donald T. and Stanley, Julian C.},
  title = {Experimental and Quasi-Experimental Designs for Research},
  year = {1966},
  publisher = {Rand McNally},
  address = {Chicago}
}

@book{shadish2002,
  author = {Shadish, William R. and Cook, Thomas D. and Campbell, Donald T.},
  title = {Experimental and Quasi-Experimental Designs for Generalized Causal Inference},
  year = {2002},
  publisher = {Houghton Mifflin},
  address = {Boston}
}

@article{linden2007,
  author = {Linden, Ariel},
  title = {Estimating the effect of regression to the mean in health management programs},
  journal = {Disease Management and Health Outcomes},
  year = {2007},
  volume = {15},
  pages = {7--12},
  doi = {https://doi.org/10.2165/00115677-200715010-00002}
}

@article{linden2016,
  author = {Linden, Ariel and Yarnold, Paul R.},
  title = {Using machine learning to identify structural breaks in single-group interrupted time series designs},
  journal = {Journal of Evaluation in Clinical Practice},
  year = {2016},
  volume = {22},
  pages = {855--859},
  doi = {https://doi.org/10.1111/jep.12544}
}

@article{linden2017a,
  author = {Linden, Ariel},
  title = {Challenges to validity in single-group interrupted time series analysis},
  journal = {Journal of Evaluation in Clinical Practice},
  year = {2017},
  volume = {23},
  pages = {413--418},
  doi = {https://doi.org/10.1111/jep.12638}
}

@article{linden2017b,
  author = {Linden, Ariel},
  title = {Persistent threats to validity in single-group interrupted time series analysis with a crossover design},
  journal = {Journal of Evaluation in Clinical Practice},
  year = {2017},
  volume = {23},
  pages = {419--425},
  doi = {https://doi.org/10.1111/jep.12668}
}

@article{linden2015,
  author = {Linden, Ariel},
  title = {Conducting interrupted time-series analysis for single- and multiple-group comparisons},
  journal = {Stata Journal},
  year = {2015},
  volume = {15},
  number = {2},
  pages = {480--500},
  doi = {https://doi.org/10.1177/1536867X1501500208}
}

@article{abadie2010,
  author = {Abadie, Alberto and Diamond, Alexis and Hainmueller, Jens},
  title = {Synthetic control methods for comparative case studies: estimating the effect of California’s tobacco control program},
  journal = {Journal of the American Statistical Association},
  year = {2010},
  volume = {105},
  number = {490},
  pages = {493--505},
  doi = {https://doi.org/10.1198/jasa.2009.ap08746}
}

@article{linden2018,
  author = {Linden, Ariel},
  title = {A matching framework to improve causal inference in interrupted time-series analysis},
  journal = {Journal of Evaluation in Clinical Practice},
  year = {2018},
  volume = {24},
  pages = {408--415},
  doi = {https://doi.org/10.1111/jep.12874}
}

@article{ryan2019,
  author = {Ryan, Andrew M. and Kontopantelis, Evangelos and Linden, Ariel and Burgess, Jr., James F.},
  title = {Now trending: Coping with non-parallel trends in difference-in-differences analysis},
  journal = {Stat Methods Med Res},
  year = {2019},
  volume = {28},
  pages = {3697--3711},
  doi = {https://doi.org/10.1177/0962280218814570}
}

@article{olden2022,
  author = {Olden, Andreas and M{\o}en, Jarle},
  title = {The triple difference estimator},
  journal = {The Econometrics Journal},
  year = {2022},
  volume = {25},
  pages = {531--553},
  doi = {https://doi.org/10.1093/ectj/utac010}
}

@article{galarraga2022,
  author = {Galarraga, J. E. and DeLia, Derek and Huang, Jing and Woodcock, Christine and Fairbanks, Richard J. and Pines, Jesse M.},
  title = {Effects of Maryland’s global budget revenue model on emergency department utilization and revisits},
  journal = {Academic Emergency Medicine},
  year = {2022},
  volume = {29},
  pages = {83--94},
  doi = {https://doi.org/10.1111/acem.14351}
}

@article{gonzales2018,
  author = {Gonzales, S. and Sommers, Benjamin D.},
  title = {Intra-ethnic coverage disparities among Latinos and the effects of health reform},
  journal = {Health Services Research},
  year = {2018},
  volume = {53},
  pages = {1373--1386},
  doi = {https://doi.org/10.1111/1475-6773.12733}
}

@book{kutner2005,
  author = {Kutner, Michael H. and Nachtsheim, Christopher J. and Neter, John and Li, William},
  title = {Applied Linear Statistical Models},
  edition = {5th},
  year = {2005},
  publisher = {McGraw-Hill Irwin},
  address = {New York}
}

@article{newey1987,
  author = {Newey, Whitney K. and West, Kenneth D.},
  title = {A simple, positive semi-definite, heteroskedasticity and autocorrelation consistent covariance matrix},
  journal = {Econometrica},
  year = {1987},
  volume = {55},
  pages = {703--708}
}

@book{box2016,
  author = {Box, George E.P. and Jenkins, Gwilym M. and Reinsel, Gregory C. and Ljung, Greta M.},
  title = {Time Series Analysis: Forecasting and Control},
  edition = {5th},
  year = {2016},
  publisher = {John Wiley \& Sons},
  address = {Hoboken}
}

@techreport{prais1954,
  author = {Prais, S. J. and Winsten, C. B.},
  title = {Trend estimators and serial correlation},
  type = {Discussion Paper},
  number = {383},
  institution = {Cowles Commission},
  year = {1954}
}

@article{cochrane1949,
  author = {Cochrane, Donald and Orcutt, Guy H.},
  title = {Application of least squares regression to relationships containing auto-correlated error terms},
  journal = {Journal of the American Statistical Association},
  year = {1949},
  volume = {44},
  pages = {32--61},
  doi = {https://doi.org/10.2307/2280349}
}

@misc{baum2013,
  author = {Baum, Christopher F. and Shaffer, Margaret E.},
  title = {ACTEST. Stata module to perform Cumby-Huizinga general test for autocorrelation in time series},
  year = {2013},
  note = {Statistical Software Components s457668, Boston College Department of Economics. Downloadable from: http://ideas.repec.org/c/boc/bocode/s457668.html}
}

@article{linden2026,
  author = {Linden, Ariel},
  title = {Power considerations for multiple-group (controlled) interrupted time series analysis: A comprehensive simulation study},
  journal = {Evaluation \& the Health Professions},
  year = {2026},
  doi = {https://doi.org/10.1177/01632787261428159}
}

@book{harvey1989,
  author = {Harvey, Andrew C.},
  title = {Forecasting, structural time series models and the Kalman filter},
  year = {1989},
  publisher = {Cambridge University Press},
  address = {Cambridge}
}

@book{enders2004,
  author = {Enders, Walter},
  title = {Applied Econometric Time Series},
  edition = {2nd},
  year = {2004},
  publisher = {John Wiley \& Sons},
  address = {New York}
}

@article{linden2006,
  author = {Linden, Ariel and Adams, John L.},
  title = {Evaluating disease management programme effectiveness: an introduction to instrumental variables},
  journal = {Journal of Evaluation in Clinical Practice},
  year = {2006},
  volume = {12},
  pages = {148--154},
  doi = {https://doi.org/10.1111/j.1365-2753.2006.00615.x}
}

@article{linden2017c,
  author = {Linden, Ariel},
  title = {A comprehensive set of postestimation measures to enrich interrupted time-series analysis},
  journal = {Stata Journal},
  year = {2017},
  volume = {17},
  pages = {73--88},
  doi = {https://doi.org/10.1177/1536867X1701700105}
}

@article{linden2022erratum,
  author = {Linden, Ariel},
  title = {Erratum: a comprehensive set of postestimation measures to enrich interrupted time-series analysis},
  journal = {Stata Journal},
  year = {2022},
  volume = {22},
  pages = {231--233},
  doi = {https://doi.org/10.1177/1536867X221083929}
}

@article{linden2005,
  author = {Linden, Ariel and Roberts, Nancy},
  title = {A user's guide to the disease management literature: recommendations for reporting and assessing program outcomes},
  journal = {American Journal of Managed Care},
  year = {2005},
  volume = {11},
  pages = {113--120}
}

\newpage
\newpage
\newpage
\begin{table}[H]
\centering
\caption{Pre-treatment and post-treatment trends, by group}
\begin{tabular}{lccc}
\toprule
Group & 
\makecell{Pre-treatment \\ (X = 0)} & 
\makecell{Post-treatment \\ (X = 1)} & 
\makecell{Pre--post \\ trend change} \\
\midrule
Control 1 (Z1 = 0, Z2 = 0) & $\beta_1$ & $\beta_1 + \beta_3$ & $\beta_3$ \\
Treatment (Z1 = 1, Z2 = 0) & $\beta_1 + \beta_5$ & $\beta_1 + \beta_3 + \beta_5 + \beta_7$ & $\beta_3 + \beta_7$ \\
Control 2 (Z2 = 1, Z1 = 0) & $\beta_1 + \beta_9$ & $\beta_1 + \beta_3 + \beta_9 + \beta_{11}$ & $\beta_3 + \beta_{11}$ \\
\bottomrule
\end{tabular}
\label{tab:trends}
\end{table}
\newpage
\begin{table}[H]
\centering
\caption{Difference-in-differences (trend change) comparisons between groups}
\begin{tabular}{lcc}
\toprule
Comparison & DiD & Simplified expression \\
\midrule
Treatment vs Control 1 & $(\beta_3 + \beta_7) - \beta_3$ & $\beta_7$ \\
Control 2 vs Control 1 & $(\beta_3 + \beta_{11}) - \beta_3$ & $\beta_{11}$ \\
Treatment vs Control 2 & $(\beta_3 + \beta_7) - (\beta_3 + \beta_{11})$ & $\beta_7 - \beta_{11}$ \\
\bottomrule
\end{tabular}
\label{tab:did_trends}
\end{table}
\newpage
\begin{table}[H]
\centering
\caption{Pre-treatment and post-treatment levels, by group}
\begin{tabular}{lccc}
\toprule
Group & 
\makecell{Pre-treatment \\ (X = 0)} & 
\makecell{Post-treatment \\ (X = 1)} & 
\makecell{Pre--post \\ level change} \\
\midrule
Control 1 (Z1 = 0, Z2 = 0) & $\beta_0$ & $\beta_0 + \beta_2$ & $\beta_2$ \\
Treatment (Z1 = 1, Z2 = 0) & $\beta_0 + \beta_4$ & $\beta_0 + \beta_2 + \beta_4 + \beta_6$ & $\beta_2 + \beta_6$ \\
Control 2 (Z2 = 1, Z1 = 0) & $\beta_0 + \beta_8$ & $\beta_0 + \beta_2 + \beta_8 + \beta_{10}$ & $\beta_2 + \beta_{10}$ \\
\bottomrule
\end{tabular}
\label{tab:levels}
\end{table}
\newpage
\begin{table}[H]
\centering
\caption{Difference-in-differences (level change) comparisons between groups}
\begin{tabular}{lcc}
\toprule
Comparison & DiD & Simplified expression \\
\midrule
Treatment vs Control 1 & $(\beta_2 + \beta_6) - \beta_2$ & $\beta_6$ \\
Control 2 vs Control 1 & $(\beta_2 + \beta_{10}) - \beta_2$ & $\beta_{10}$ \\
Treatment vs Control 2 & $(\beta_2 + \beta_6) - (\beta_2 + \beta_{10})$ & $\beta_6 - \beta_{10}$ \\
\bottomrule
\end{tabular}
\label{tab:did_levels}
\end{table}
\newpage
\begin{table}[H]
\centering
\caption{Regression results for the DDD-ITSA model}
\begin{threeparttable}
\begin{tabular}{ll
                 S[table-format=3.2]
                 S[table-format=1.2]
                 S[table-format=2.2]
                 S[table-format=1.3]
                 S[table-format=3.2]
                 S[table-format=3.2]}
\toprule
\multicolumn{2}{l}{Label} & {Coeff.} & {Std Err.} & {Z} & {$P>|z|$} & {95\% LCL} & {95\% UCL} \\
\midrule
$\beta_0$  & (const)         & 126.40 & 4.58 & 27.57 & 0.001 & 117.41 & 135.38 \\
$\beta_1$  & (T)             & -1.43  & 0.43 & -3.35 & 0.001 & -2.27 & -0.59 \\
$\beta_2$  & (X)             & -11.43 & 4.46 & -2.56 & 0.010 & -20.17 & -2.68 \\
$\beta_3$  & (X$\times$T)   & 0.58   & 0.61 & 0.95  & 0.345 & -0.62 & 1.78 \\
$\beta_4$  & (Z1)            & 5.83   & 6.20 & 0.94  & 0.347 & -6.33 & 17.98 \\
$\beta_5$  & (Z1$\times$T)  & -0.35  & 0.57 & -0.61 & 0.539 & -1.47 & 0.77 \\
$\beta_6$  & (Z1$\times$X)  & -8.63  & 6.44 & -1.34 & 0.180 & -21.25 & 3.98 \\
$\beta_7$  & (Z1$\times$X$\times$T) & -2.07  & 0.75 & -2.77 & 0.006 & -3.54 & -0.61 \\
$\beta_8$  & (Z2)            & 12.29  & 6.84 & 1.80  & 0.072 & -1.11 & 25.70 \\
$\beta_9$  & (Z2$\times$T)  & -0.10  & 0.69 & -0.15 & 0.880 & -1.46 & 1.25 \\
$\beta_{10}$ & (Z2$\times$X) & -6.48 & 7.75 & -0.84 & 0.403 & -21.67 & 8.70 \\
$\beta_{11}$ & (Z2$\times$X$\times$T) & -0.31 & 0.88 & -0.35 & 0.723 & -2.03 & 1.41 \\
\bottomrule
\end{tabular}
\begin{tablenotes}
\small
\item Notes: Labels correspond to the coefficient labels in Equation~\eqref{eq:model}; standard errors are Newey--West AR(1) adjusted.
\end{tablenotes}
\end{threeparttable}
\label{tab:regression}
\end{table}
\newpage
\FloatBarrier

\newpage
\section*{Abbreviations}

ITSA: Interrupted time series analysis. \\
MG-ITSA: Multiple-group interrupted time series analysis. \\
SG-ITSA: Single-group interrupted time series analysis. \\
DiD: Difference-in-differences. \\
DDD-ITSA: Triple difference interrupted time series analysis. \\
CI: Confidence interval. \\
GBR: Maryland's global budget revenue (GBR) model. \\
ACA: Affordable Care Act.

\section*{Declarations}

\begin{description}
\item[\textbf{Ethics approval and consent to participate}] Not applicable.
\item[\textbf{Consent for publication}] Not applicable.
\item[\textbf{Availability of data and materials}] All the code used in the Example is provided in the Appendix at the end of this paper.
\item[\textbf{Competing interests}] The author declares no competing interests.
\item[\textbf{Funding}] There was no funding associated with this work.
\end{description}

\section*{Authors' contributions}

AL conceived the study and its design, conducted all analyses, wrote the manuscript and takes public responsibility for its content.

\section*{Acknowledgements}

None.

% -------------------------
% Appendix & misc
% -------------------------
\newpage
\section*{Appendix}

Stata code used in the Proposition 99 example

\begin{verbatim}
// Download the most recent version of -itsa-
ssc install itsa, replace

// Example
use "cigsales.dta", replace

// Declare data as time series
tsset state year

* estimate ITSA model in which California is specified as the 
* treatment unit (#3 in the data), the treatment period is set 
* to 1989, the lag is set to 1 (based on actest), the primary 
* control IDs are 8 and 19 (for Idaho and Montana) and the secondary 
* control is Colorado (4). We specify that the output should 
* include post-treatment trends for all groups, and request that 
* a figure be produced

itsa cigsale, treatid(3) trperiod(1989) lag(1) replace ///
contid(8 19) contid2(4) posttrend figure

* baseline level treatment vs control 1 (beta4)
lincom  _b[_z1]

* baseline trend treatment vs control 1 (beta5)
lincom _b[_z1_t]

* baseline level control 2 vs control 1 (beta8)
lincom  _b[_z2]

* baseline trend control 2 vs control 1 (beta9)
lincom _b[_z2_t]

* baseline level treatment vs control 2 (beta4 - beta8)
lincom  _b[_z1] - _b[_z2]

* baseline trend treatment vs control 2 (beta5 - beta9)
lincom _b[_z1_t] - _b[_z2_t]    

* DD trends treatment vs control 1 (beta7)
lincom _b[_z1_x_t1989]

* DD trends treatment vs control 2 (beta7 - beta11)
lincom _b[_z1_x_t1989] - _b[_z2_x_t1989]

* DD trends control 1 vs control 2 (beta11)
lincom _b[_z2_x_t1989]

* DDD-ITSA (beta7 - beta11)
lincom _b[_z1_x_t1989] - _b[_z2_x_t1989]

* Sensitivity analysis: group-specific quadratic time trends 
* entered as covariates

* generate quadratic time terms for each group
gen t2 = (year - 1970)^2
gen z1_t2 = (state == 3) * t2
gen z2_t2 = (state == 4) * t2

itsa cigsale t2 z1_t2 z2_t2, treatid(3) trperiod(1989) lag(1) ///
replace contid(8 19) contid2(4)

* common post-intervention level change (beta2)
lincom _b[_x1989]

* DDD level change treatment vs control 2 (beta6 - beta10)
lincom _b[_z1_x1989] - _b[_z2_x1989]

* DDD-ITSA (beta7 - beta11)
lincom _b[_z1_x_t1989] - _b[_z2_x_t1989]
\end{verbatim}

\end{document}